\documentclass{elsart}
\usepackage{graphicx}
\usepackage{dcolumn}
\usepackage{cite}

\begin{document}

\begin{frontmatter}
\title{Accurate calculation of the solutions to the Thomas-Fermi equations}

\author{Paolo Amore\thanksref{PA}}

\address{Facultad de Ciencias, CUICBAS, Universidad de Colima, \\
Bernal D\'{\i}az del Castillo 340, Colima, Colima, Mexico
}
\author{John P. Boyd\thanksref{JB}}

\address{Department of Atmospheric, Oceanic \& Space
Science  \\University of Michigan, 2455 Hayward Avenue, Ann Arbor
MI 48109}

\author{Francisco M. Fern\'{a}ndez\thanksref{FMF}}

\address{INIFTA (UNLP, CCT La Plata-CONICET), Divisi\'{o}n Qu\'{i}mica Te\'{o}rica,\\
Blvd. 113 y 64 (S/N), Sucursal 4, Casilla de Correo 16,\\
1900 La Plata, Argentina}

\thanks[PA]{e-mail: paolo.amore@gmail.com}
\thanks[JB]{e-mail: jpboyd@umich.edu; \\ http://www.engin.umich.edu:/$\sim$ jpboyd}
\thanks[FMF]{e-mail: fernande@quimica.unlp.edu.ar}

\begin{abstract}
We obtain highly accurate solutions to the Thomas-Fermi equations
for atoms and atoms in very strong magnetic fields. We apply the
Pad\'e-Hankel method, numerical integration, power series with
Pad\'e and Hermite-Pad\'e approximants and Chebyshev polynomials.
Both the slope at origin and the location of the right boundary in
the magnetic-field case are given with unprecedented accuracy.
\end{abstract}

\end{frontmatter}

\section{Introduction}

\label{sec:intro}

The Thomas-Fermi model is one of the simplest approaches to the study of the
potential and charge densities in a variety of systems, like, for example,
atoms\cite{B30,CM50,KMNU55,M57,T72, M83}, molecules\cite{M52,M57}, atoms in
strong magnetic fields\cite{BCR74,TY78,MT79,M83,HGM83}, metals and crystals%
\cite{SK35,UT55} and dense plasmas\cite{YK89}. For this reason there has
been great interest in the accurate calculation of the solution to the
Thomas-Fermi equation\cite
{B30,BC31,KMNU55,FMT49,T72,PP87,FO90,AF07,AF11,F08,F11,AB11,B12}. In
particular the accurate results obtained by Kobayashi et al\cite{KMNU55} by
numerical integration are commonly chosen as benchmark data for testing
other approaches. The even more accurate results of Rijnierse do not appear
to be so well known, probably because they do not appear to have been
published and are only quoted in the book by Torrens\cite{T72}.

The behaviour of the solution to the nonlinear Thomas-Fermi
equation depends on the slope at the origin. The critical slope at
the origin suitable for neutral atoms is of particular interest
and has been estimated by many authors (see, for example,
Kobayashi et al\cite{KMNU55}). Particularly accurate results for
this critical slope were obtained by Amore and
Fern\'{a}ndez\cite{AF07,AF11} and later by
Fern\'{a}ndez\cite{F08,F11} using the Pad\'{e}-Hankel method
(PHM). Abbasbandy and Bervillier\cite{AB11} considerably improved
this estimate by means of a judicious analytic continuation of the
expansion of the solution about the origin and Boyd\cite {B12}
reported an even more accurate result obtained by means of a
rational Chebyshev series. Other authors have also resorted to
Pad\'e approximants in order to approximate the solution to the
Thomas-Fermi equation\cite{T91,EFGP99}. It has been shown that a
method due to Majorama is suitable for obtaining a semi-analytical
series solution to the Thomas-Fermi equation in terms of only one
quadrature\cite{E02}.

The analytic properties of the solution of the Thomas-Fermi equation under
different boundary conditions (in addition to the physically relevant ones)
are also of great interest and have been studied by several authors (see,
for example, Hille\cite{H69,H70} and the references therein).

The purpose of this paper is twofold. First, we want to stress the
different behaviour of the solution of the Thomas-Fermi equation
for atoms and atoms in strong magnetic fields that was overlooked
in a recent application of the PHM\cite{F11}. More precisely, in
his application of the PHM Fern\'{a}ndez\cite{F11} assumed the
incorrect asymptotic behaviour at infinity suggested by Banerjee
et al\cite{BCR74}. However, since the PHM does not take into
account the second (outer) boundary condition explicitly
Fern\'{a}ndez obtained an accurate slope at the origin. The PHM is
based on the Riccati-Pad\'e method that was developed to obtain
bound states and resonances of separable quantum-mechanical
problems\cite{FMT89,F92,FG93,F95,F95b,F95c,F96,F96b,F97}.

Our second goal is to show that the available computer-algebra
software enable one to solve the Thomas-Fermi equations (and,
certainly, other nonlinear equations as well) with great accuracy.
We want to provide sufficiently accurate solutions that may be
used as benchmark data for testing future analytical or numerical
methods.

Boyd\cite{B12} obtained the most accurate critical slope at the origin of
the solution to the Thomas-Fermi equation for neutral atoms by means of
rational Chebyshev series. In this paper we try the Chebyshev polynomials on
the equation for neutral atoms in very strong magnetic fields.

In Section \ref{sec:expansions} we outline the expansions of the
solution to the Thomas-Fermi equation about origin, at infinity,
about poles and zeroes/branch points. Such expansions have already
been discussed by other authors and also used as aids for
obtaining approximate analytical solutions as well as accurate
numerical results\cite
{B30,FMT49,M52,KMNU55,H69,H70,T72,M83,PP87,FO90,T91,CM50,AB11}.
The main purpose of this section is to stress the difference
between the Thomas-Fermi equations for isolated atoms and atoms in
strong magnetic fields. In section \ref{sec:PHM} we describe the
accurate results obtained by means of the PHM
and from straightforward integration of the differential equations. In Sec.~%
\ref{sec:Pow_ser} we discuss the application of two power series
and their Pad\'{e} and Hermite-Pad\'{e} approximants to the
Thomas-Fermi equation for a neutral atom in a magnetic field. In
Sec.~\ref{sec:Cheb_1} we apply Chebyshev polynomials to the same
problem and in Sec.~\ref{sec:Cheb_2} we consider analytical
results based such polynomials of small degree. Finally, in
Sec.~\ref{sec:conclusions} we draw conclusions.

\section{Expansions for the solutions to the Thomas-Fermi equations}

\label{sec:expansions}

In order to facilitate the discussion and to make this paper clearer in this
section we summarize the well known expansions of the Thomas-Fermi equations
about some characteristic points. As indicated above, such expansions are
well known and have been widely used by other authors for several different
purposes\cite{B30,FMT49,M52,KMNU55,H69,H70,T72,M83,PP87,FO90,T91,CM50,AB11}.

In this paper we restrict ourselves to the simplest cases. The first one is
the Thomas-Fermi equation for an atom
\begin{eqnarray}
u^{\prime \prime }(x) &=&\frac{u(x)^{3/2}}{x^{1/2}}  \nonumber \\
u(0) &=&1  \label{eq:TF_eq}
\end{eqnarray}
It can be expanded about origin as
\begin{equation}
u=1+ax+\frac{4x^{3/2}}{3}+\frac{2ax^{5/2}}{5}+\frac{x^{3}}{3}+\ldots
\label{eq:TF_origin}
\end{equation}
where $a=u^{\prime }(0)<0$ is the unknown slope at the origin.

There is a critical slope $u_{0}^{\prime }$ and the behaviour of the
solution depends on the relation between $u^{\prime }(0)$ and $u_{0}^{\prime
}$. If $u^{\prime }(0)$ $<$ $u_{0}^{\prime }$ the solution vanishes at a
movable branch point\cite{H70} $x=x_{q}$ around which it behaves as
\begin{equation}
u=bs+b^{3/2}x_{q}^{3/2}\left( \frac{4s^{7/2}}{35}+\frac{2s^{9/2}}{63}+\frac{%
s^{11/2}}{66}+\ldots \right)  \label{eq:TF_zero}
\end{equation}
where $s=(x_{q}-x)/x_{q}$ and $b$ is a constant.

If $u^{\prime }(0)$ $>$ $u_{0}^{\prime }$ the solution decreases and reaches
a minimum at $x=x_{m}$ about which it behaves as
\begin{equation}
u=a_{0}+(a_{0}x_{m})^{3/2}\left[ \frac{s^{2}}{2}+\frac{s^{3}}{12}+\frac{%
\left( 2\sqrt{a_{0}}x_{m}^{3/2}+1\right) s^{4}}{32}+\frac{\left( 8\sqrt{a_{0}%
}x_{m}^{3/2}+5\right) s^{5}}{320}+\ldots \right]  \label{eq:TF_min}
\end{equation}
where $s=(x_{m}-x)/x_{m}$ and $a_{0}$ is a positive constant.

After this minimum the solution increases and tends to infinity because of a
movable singularity at $x=x_{s}$ about which it behaves as
\begin{equation}
u=\frac{1}{x_{s}^{3}}\left( \frac{400}{s^{4}}-\frac{2000}{9s^{3}}-\frac{2000%
}{81s^{2}}-\frac{10000}{729s}-\frac{3189895}{177147}+\ldots \right)
\label{eq:TF_pole}
\end{equation}
where $s=(x_{s}-x)/x_{s}$. Abbasbandy and Bervillier\cite{AB11} derived an
alternative and more detailed expansion in terms of the variable $z=x^{1/2}$
for the function $g(z)=\sqrt{u(z^{2})}$.

It is also well known that $x_{q}\rightarrow \infty $ as $u^{\prime
}(0)\rightarrow u_{0}^{\prime }$ from the left and $x_{m},x_{s}\rightarrow
\infty $ as $u^{\prime }(0)\rightarrow u_{0}^{\prime }$ from the right. At
this limit the solution tends monotonically to zero according to
\begin{equation}
u=\frac{144}{x^{3}}+d\,\,x^{(1-\sqrt{73})/2}+\frac{7\sqrt{73}+67}{29184}%
d^{2}x^{4-\sqrt{73}}+\ldots  \label{eq:TF_infinity}
\end{equation}
where $d$ is a constant. There have been reasonably successful attempts at
matching the expansions (\ref{eq:TF_origin}) and (\ref{eq:TF_infinity}) by
means of appropriate nonlinear transformations of the independent variable%
\cite{PP87,FO90}.

From a physical point of view the zero at $x=x_{q}$ (see the
expansion (\ref {eq:TF_zero})) is the radio of an atom if
\begin{equation}
-x_{q}u^{\prime }(x_{q})=1-\frac{N}{Z}=q  \label{eq:BC_xq}
\end{equation}
where $N$ is the number of electrons, $Z$ is the atomic number and $q$ is
the degree of ionization (note that $N<Z$). For a neutral atom ($q=0$) the
boundary condition $u(x_{0})=u^{\prime }(x_{0})=0$ takes place at $%
x_{0}=\infty $ as indicated above for the case $u^{\prime }(0)=u_{0}^{\prime
}$.

Abbasbandy and Bervillier\cite{AB11} argued that the PHM applies
successfully to this problem because the Hankel condition sends the movable
singularity at $x=x_{s}$ to infinity.

The Thomas-Fermi equation for an atom in a strong magnetic field is
(Tomishina and Yonei\cite{TY78} proposed a somewhat more realistic model
that we do not discuss here)
\begin{eqnarray}
u^{\prime \prime }(x) &=&\sqrt{xu(x)}  \nonumber \\
u(0) &=&1  \label{eq:TFF_eq}
\end{eqnarray}
In this case the expansion about the origin is given by

\begin{equation}
u=1+ax+\frac{4x^{5/2}}{15}+\frac{2ax^{7/2}}{35}+\ldots  \label{eq:TFF_origin}
\end{equation}
where $a=u^{\prime }(0)$. As in the preceding case the behaviour of the
solution depends on this slope at the origin that also exhibits a critical
value $u_{0}^{\prime }$.

If $u^{\prime }(0)<u_{0}^{\prime }$ the solution vanishes at a movable
branch point $x=x_{q}$ according to
\begin{equation}
u=bs+\frac{4\sqrt{b\,x_{1}^{5}}s^{5/2}}{15}-\frac{2\sqrt{b\,x_{1}^{5}}s^{7/2}%
}{35}+\ldots  \label{eq:TFF_zero}
\end{equation}
where $s=(x_{q}-x)/x_{q}$ and $b$ is a constant.

If $u^{\prime }(0)>u_{0}^{\prime }$ the solution exhibits a minimum at $%
x=x_{m}$ around which it behaves as
\begin{equation}
u=a_{0}+\frac{\sqrt{a_{0}x_{m}^{5}}}{2}s^{2}-\frac{\sqrt{a_{0}x_{m}^{5}}}{12}%
s^{3}+\frac{2x_{m}^{5}-\sqrt{a_{0}x_{m}^{5}}}{96}s^{4}-\frac{3\sqrt{%
a_{0}x_{m}^{5}}+8x_{m}^{5}}{960}s^{5}+\ldots  \label{eq:TFF_min}
\end{equation}
where $s=(x_{m}-x)/x_{m}$ and $a_{0}$ is a constant.

The main difference between this equation and the preceding one is that in
this case the pole is located at infinity. For large values of the
coordinate the solution behaves as
\begin{equation}
u=\frac{x^{5}}{400}+c_{0}x^{\left( \sqrt{41}+1\right) /2}+c_{0}^{2}\left(
\frac{25\sqrt{41}}{2}+\frac{425}{6}\right) x^{\sqrt{41}-4}+\ldots
\label{eq:TFF_infinity}
\end{equation}
where $c_{0}$ is a constant.

When $u^{\prime }(0)=u_{0}^{\prime }$ the solution and its first derivative
vanishes at $x=x_{0}$ according to the expansion
\begin{equation}
u=x_{0}^{5}s^{4}\left( \frac{1}{144}-\frac{s}{336}-\frac{s^{2}}{7056}-\frac{%
s^{3}}{16464}\ldots \right)  \label{eq:TFF_crit}
\end{equation}
where $s=(x_{0}-x)/x_{0}$. The location of the minimum $x_{m}$ approaches $%
x_{0}$ from below as $u^{\prime }(0)$ approaches $u_{0}^{\prime }$ from
above. The zero/branch point $x_{q}$ also approaches $x_{0}$ from below as $%
u^{\prime }(0)$ approaches $u_{0}^{\prime }$ from below. However, the
function is analytic at $x=x_{0}$ when $u^{\prime }(0)=u_{0}^{\prime }$ as
shown in Eq.~(\ref{eq:TFF_crit}). The solution with the critical slope at
the origin also tends to infinity as $x\rightarrow \infty $ according to
equation (\ref{eq:TFF_infinity}). According to Banerjee et al\cite{BCR74}
the universal solution corresponding to neutral atoms ($N=Z$) satisfies $%
u(x_{0})=0$ and $u^{\prime }(x_{0})=0$ at $x_{0}=\infty $. Based
on this conjecture Fern\'{a}ndez\cite{F11} applied the PHM and
obtained a somewhat more accurate value of $u_{0}^{\prime }$.
However Hill et al\cite{HGM83} showed that $x_{0}$ is finite and
can be related to $x_{q}$ by means of a perturbation expansion of
the form.
\begin{equation}
x_{q}=x_{0}-(24q/x_{0}^{2})^{1/3}+\ldots  \label{eq:x_q-->x_0}
\end{equation}
which clearly shows that $x_{q}$ approaches $x_{0}$ from below as $%
q\rightarrow 0$ (and $u^{\prime }(0)\rightarrow u_{0}^{\prime }$ from
below). They confirmed this result by numerical integration of Eq.~(\ref
{eq:TFF_eq}).

In the two cases discussed above the PHM yields the correct critical slope
at the origin disregarding the second boundary condition. In the first
example both $u(x)$ and $u^{\prime }(x)$ vanish as $x\rightarrow \infty $,
while in the second example they vanish at a finite value $x_{0}$ of the
independent variable. Abbasbandy and Bervillier\cite{AB11} suggested that
the success of the PHM is based on ``forcing the localization at infinity of
a movable singularity (when it exists)''. They also stated that ``if the
second boundary is located at infinity, the PHM has a particular
significance''. The success of the PHM for the Thomas-Fermi equation (\ref
{eq:TFF_eq}) shows that the approach is also suitable when the second
boundary condition takes place at a finite point. In this case the movable
singularity pushed to infinity may be the zero/branch point $x_{q}$
discussed above (Eq.~(\ref{eq:TFF_zero})). It was shown that $x_{q}$
approaches $x_{0}$ as $u^{\prime }(0)$ approaches the critical slope but it
may jump to infinity when $u^{\prime }(0)=u_{0}^{\prime }$ leaving the
solution analytic at $x_{0}$ as discussed above. We cannot prove this
conjecture rigorously but we believe that it sounds plausible.

\section{PHM and numerical integration}

\label{sec:PHM}

We first review the main points of the PHM. Following Amore and
Fern\'{a}ndez\cite{AF07,AF11,F08,F11} we choose the new variables
$x=t^{2}$ and $v(t)=\sqrt{u(t^{2})}$. The function $v(t)$ can be
expanded in a Taylor series
\begin{equation}
v(t)=\sum_{j=0}^{\infty }v_{j}t^{j}
\end{equation}
where the coefficients $v_{j}$, $j\geq 4$ depend on
$v_{2}=u_{0}^{\prime }/2$.

We construct the Hankel determinants $H_{D}^{d}=\left| v_{i+j+d+1}\right|
_{i,j=0,1,\ldots D-1}$ that depend on the unknown slope at the origin $%
u_{0}^{\prime }$ and obtain sequences of roots $u_{0}^{\prime `}[D,d]$ of $%
H_{D}^{d}=0$, $D=2,3,\ldots $, ($d=0,1,\ldots $ fixed) that converge towards
the critical slope.

We carried out PHM calculations for values of $D$ greater than
those used before\cite{F11} and also numerical integration based
commands built in Mathematica together with the bisection method.
We describe the results in what follows.

There are far too many results for the critical slope of the solution to the
Thomas-Fermi equation for neutral atoms. Table~\ref{tab:TF_crit} just shows
the most accurate ones. Present PHM result was estimated from sequences of
roots of the Hankel determinants with $D\leq 60$ and $d=3$.

The Thomas-Fermi equation for a neutral atom in a strong magnetic field has
not been so widely studied. Table~\ref{tab:TFF_crit} shows the available
critical slopes at origin. Present PHM result was estimated by comparing
results with $D\leq 60$ and $d=1,\,3$. The PHM exhibits a much greater rate
of convergence for this problem.

In order to provide benchmark data for testing other approaches in
the future we have calculated $u(x)$ and $u^{\prime }(x)$ for
Eq.~(\ref{eq:TF_eq}) and $u(x)$ for Eq.~(\ref{eq:TFF_eq}) as
accurately as possible using straightforward numerical
integration. In principle, we can resort to the analytical
behaviour of the solutions described in section
\ref{sec:expansions} in order to bracket the slope at origin
$u^{\prime}_0$ with any desired accuracy. However, in the present
case we have remarkably accurate values of the critical slope at
origin obtained by other methods and, therefore, we proceeded in a
different way that we describe in what follows.

A sufficiently accurate value of the slope at origin for the
Thomas-Fermi equation obtained in this paper is
$u'(0)=-1.588071022611375312718684508$. We calculated the values
of $u(x)$ and $u^{\prime}(x)$ by means of the command {\rm
NSDolve} built in Mathematica with different accuracy choices to
test the precision of the numerical results. We tried three sets
of parameters:
\begin{itemize}
\item Case I:  WorkingPrecision=300; PrecisionGoal=20;AccuracyGoal=20;
\item Case II:  WorkingPrecision=1000; PrecisionGoal=50;AccuracyGoal=50;
\item Case III:  WorkingPrecision=1000; PrecisionGoal=60;AccuracyGoal=60;
\end{itemize}
and carried out the numerical integration through the interval
$(0,1000)$.

In addition to the straightforward comparison of the results for
those three sets it is also necessary to determine the effect of
the error due to the approximate choice of $u^{\prime}_0$. With
this purpose in mind we also carried out the calculations with
slightly modified critical slopes:
\begin{itemize}
\item $u_L'(0)=-1.588071022611375312718684509$

\item $u_R'(0)=-1.588071022611375312718684507$
\end{itemize}
The maximum difference between the values of $u(x)$ obtained with
sets I and III at all the chosen points was found to be of order
$10^{-14}$, while such difference for sets II and III was $2
\times 10^{-20}$. On the other hand, even considering less
accurate set I, we found that the maximum difference between
values of $u(x)$ calculated with  $u_L'(0)$ and $u_R'(0)$ was of
the order of $10^{-17}$. Thus we conclude that set I yields
sufficiently accurate results with the critical slope at origin
shown above.

A sample of the results is shown in tables \ref {tab:TF},
\ref{tab:TFD} and \ref{tab:TFF}, and a considerably wider range of
values is available elsewhere\cite{ABF12}. From the numerical
analysis outlined above we are confident that all the digits in
those tables are exact.

\section{Power-series approaches}

\label{sec:Pow_ser}

\subsection{Power-series for the original equation}

When $u^{\prime }(0)=u_{0}^{\prime }$ the function is analytic at $x=x_{0}$
and numerical experimentation suggests that the Taylor series (\ref
{eq:TFF_crit}) converges over the whole domain of interest, even at the left
endpoint $x=0$. It is convenient to rescale the coordinate by defining $%
s=(x_{0}-x)/x_{0}$ so that $x=0$ corresponds to $s=-1$. Therefore we may try
to calculate the parameters $u_{0}^{\prime }$ and $x_{0}$ by means of the
sequences of partial sums
\begin{equation}
u_{M}(x)=x_{0}^{5}s^{4}\sum_{j=0}^{M}c_{j}s^{j},\,s=\frac{x_{0}-x}{x_{0}}
\label{eq:TFF_crit_N}
\end{equation}
and the conditions $u_{M}(0)=1$, $u_{M}^{\prime }(0)=u_{0}^{\prime }$. Since
the singularity at $s=-1$ ($x=0$) is proportional to $x^{5/2}$ (see Eq.~(\ref
{eq:TFF_origin})) the coefficients $c_{j}$ decrease as $O(j^{-7/2})$ and the
error of the $M-th$ partial sum falls proportionally to $M^{-5/2}$. For
example, in this way we obtain $x_{0}\approx 3.06882$ that is correct for
five digits versus the actual value $x_{0}\approx 3.068857$. It is
remarkable that a power series is able to yield any accuracy at all for a
nonlinear, singular boundary value problem.

\subsection{Quadratic Pad\'{e} approximants}

Accelerating the convergence by applying ordinary Pad\'{e} approximations
produced no improvement because the function $u$ has a branch point at the
left boundary, precisely where we are summing the series to approximate $%
u^{\prime }(0)$. However, the so-called ``Hermite-Pad\'{e}'' or ``Shafer''
approximation is much more successful.

The quadratic Shafer approximant $u[K/L/M](s)$ is defined to be the solution
of the quadratic equation \cite{SG98,S74,BN02,B99}
\begin{equation}
P(s)\,(u[K/L/M])^{2}+Q(s)\,u[K/L/M]\,+\,R(s)=0
\end{equation}
where the polynomials $P$, $Q$ and $R$ are of degrees $K$, $L$ and $M$,
respectively. These polynomials are chosen so that the power series
expansion of $u[K/L/M](s)$ agrees with that of $u(s)$ through the first $%
N=K+L+M+1$ terms. The constant terms in $P$ and $Q$ can be set arbitrarily
to one without loss of generality since these choices do not alter the roots
of the equation, so the total number of degrees of freedom is $N=K+L+M+1$.
As true for ordinary Pad\'{e} approximants, the coefficients of the
polynomials can be computed by solving a matrix equation and the most
accurate approximations are obtained by choosing the polynomials to be of
equal degree, so-called ``diagonal'' approximants. Because the power series
begins with $s^{4}$, the method was applied to $\tilde{u}\equiv
u(x[s])/(x_0^{5}s^{4}$ in the coordinate $s$; the eigenparameter is then
estimated from
\begin{equation}
x_0 \approx \tilde{u}(s=-1)^{-1/5}
\end{equation}

\begin{figure}[b]
\centerline{\includegraphics[scale=0.5]{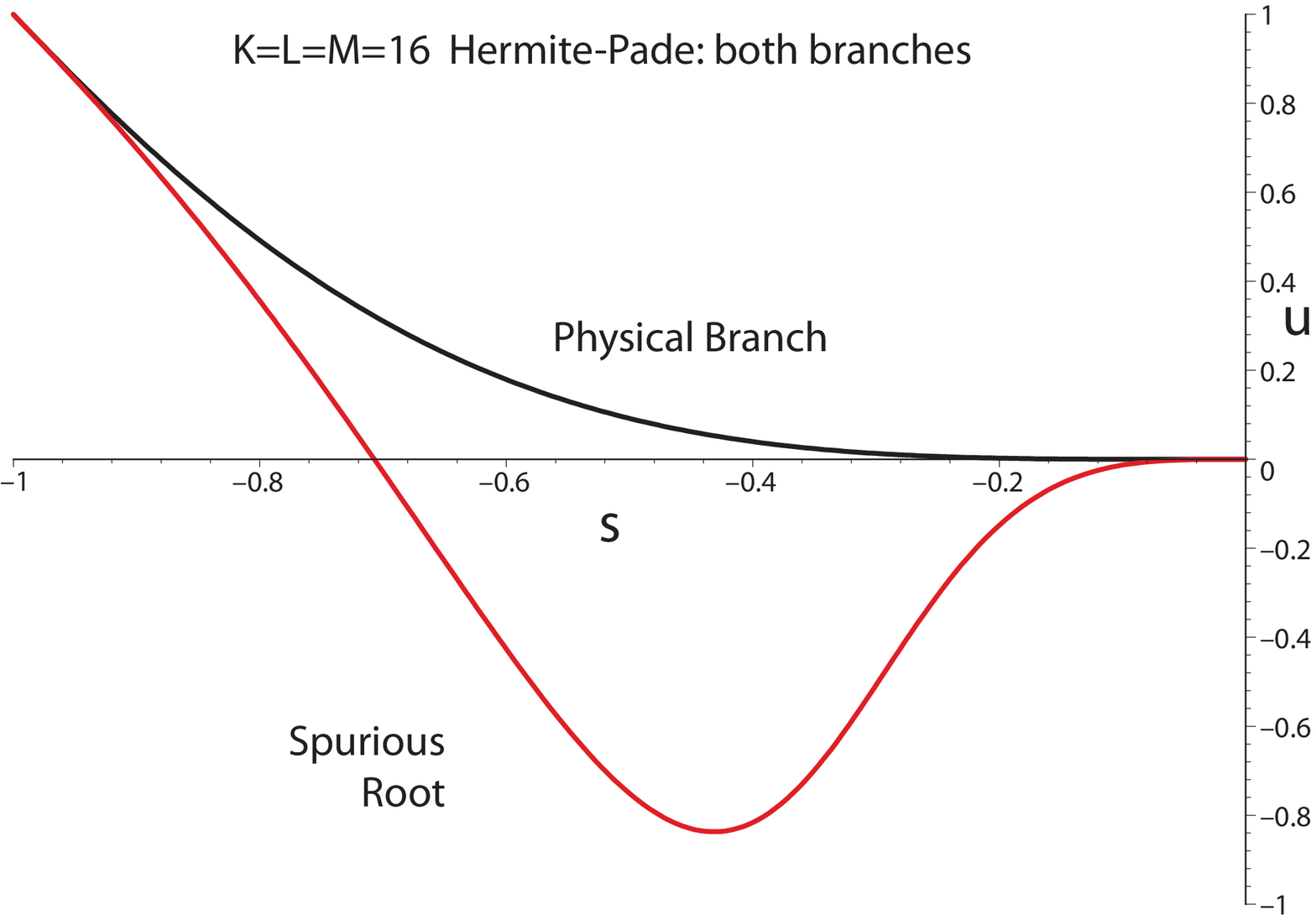}}
\caption{The $K=L=M=16$ Hermite-Pad\'e approximations to $u$ in the
coordinate $s = x/x_0 - 1$.}
\label{FigTFMagHermitePadeKLM16_bothroots}
\end{figure}

\begin{figure}[b]
\centerline{\includegraphics[scale=0.5]{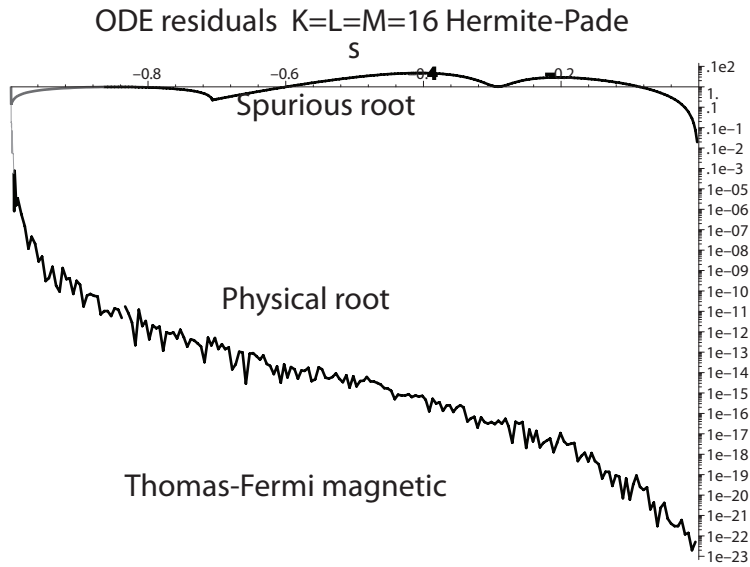}}
\caption{The $K=L=M=16$ Hermite-Pad\'e ODE residuals, that is, the result of
substituting the approximation into the Thomas-Fermi differential equation. }
\label{FigTFMagHermitePadeKLM16_bothresid}
\end{figure}

The quadratic equation actually yields \emph{two} approximations:
\begin{equation}
~ u^{\pm}[K/L/M](s) = \frac{- Q(s) \,\pm \sqrt {Q(s) ^ {2} - 4 P (s)\, R(s)}%
} {2 \, P (s)}  \label{Eqqqq}
\end{equation}
as illustrated for $K=L=M=16$ in Fig.~\ref
{FigTFMagHermitePadeKLM16_bothroots}; the root obtained from applying the
minus sign in front of the radical is spuriously negative on much of the
interval $s \in [-1, 0] $. Substituting $u=x_0^{5} s^{4} u^{\pm}[K/L/M]$
into the differential equation yields the ``residual function, $%
\rho^{\pm}=u_{ss}/x_0^{2} - \sqrt{ x_0 (s+1) u}$. Plotting the residual
functions as in Fig.~\ref{FigTFMagHermitePadeKLM16_bothresid} shows that the
``physical'' root is the plus sign in (\ref{Eqqqq}); the residual function
is tiny for the plus root but $O (1) $ for the negative root. One warning is
that, like ordinary Pad\'{e} approximants, Hermite-Pad\'{e} computations are
rather ill-conditioned. We therefore employed Maple so that much of the work
was done in exact rational arithmetic, and floating points portions were
calculated using 60 decimal digits of precision; we did not investigate
ill-conditioning.

The first surprise is that \emph{both} solution branches of the
Hermite-Pad\' {e} \emph{converge} to almost the same values as $s
\leftrightarrow -1 $. We therefore list the errors of the diagonal
Shafer approximants for $x_0$ from both solutions in
Table~\ref{Tab1eHermPad}. The table also gives the error of the
power series up to order $N $ from whence the Hermite-Pad\'{e}
approximations were obtained. We noticed that the errors of the
two branches
are roughly equal and opposite; we therefore also list the error in the\emph{%
average} of the two roots of the quadratic. Each of the roots of the
quadratic is a much better approximation than the power series.

In other applications of Hermite-Pad\'{e} approximations, the roots converge
to separate modes, such as one root to the ground state eigenvalue and the
other to the first excited state. Here, however, both roots converge to the
unique eigenparameter and their average is extraordinarily accurate. We are
unaware of another profitable use of averaging in Hermite-Pad\'{e}
approximations.

\begin{figure}[b]
\centerline{\includegraphics[scale=0.5]{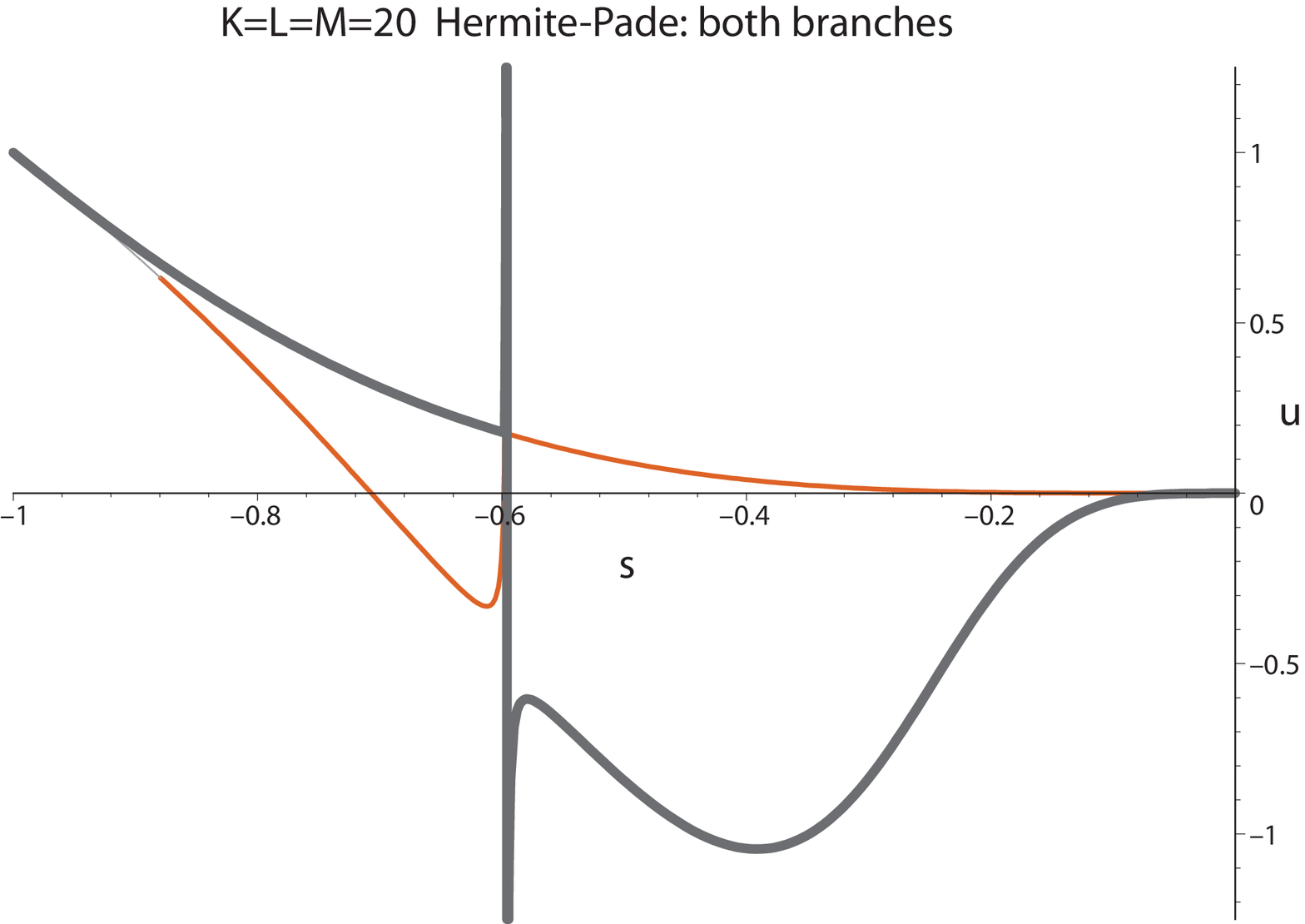}}
\caption{The $K=L=M=20$ Hermite-Pad\'e approximations to $u$ in
the coordinate $s = z/x_0 - 1$. }
\label{FigTFMagHermitePadeKLM20_bothroots}
\end{figure}

\begin{figure}[b]
\centerline{\includegraphics[scale=0.5]{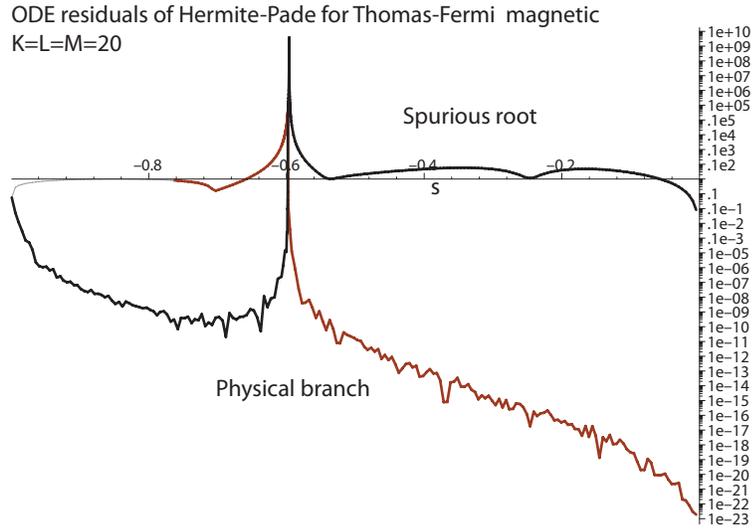}}
\caption{The $K=L=M=20$ Hermite-Pad\'e ODE residuals in the
coordinate $s = z/x_0 - 1$. }
\label{FigTFMagHermitePadeKLM20_bothresid}
\end{figure}

\begin{figure}[b]
\centerline{\includegraphics[scale=0.5]{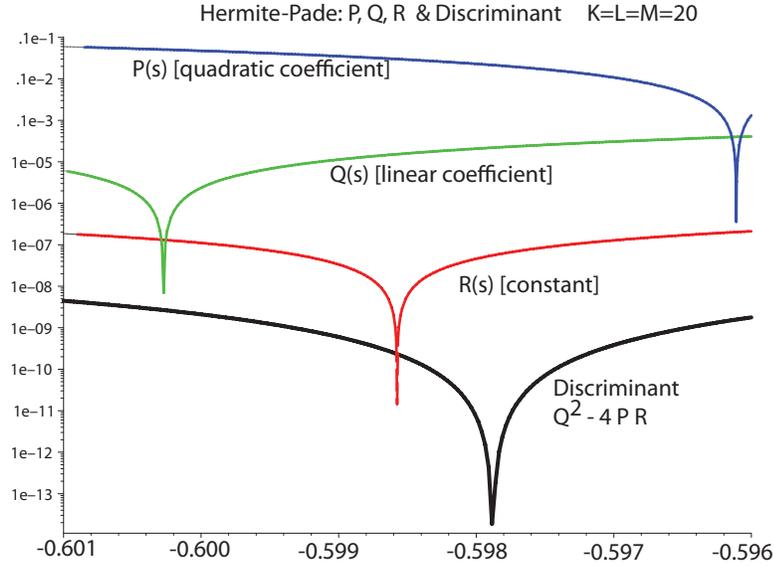}%
}
\caption{The $K=L=M=20$ Hermite-Pad\'e approximant: Plots of the coefficient
polynomials in the quadratics equation $P (f[K/L/M])^{2} + Q \, f[K/L/M] \,
+ \, R = 0 $ and also the ``discriminant'', $D \equiv Q ^ {2} - 4 P R $,
which is the argument of the square root in the solutions to the quadratic. }
\label{FigTFMagHermitePadeKLM20_PQRDiscrim}
\end{figure}

The second surprise is that when $K \geq 17 $, the diagonal approximant
develops a spurious singularity in the middle of the spatial interval for
both branches as illustrated in Fig.~\ref{FigTFMagHermitePadeKLM20_bothroots}%
. Except in a narrow neighborhood of the singularity, the residual function
for one branch is very tiny as illustrated in Fig.~\ref
{FigTFMagHermitePadeKLM20_bothresid}. It has long been known that ordinary
Pad\'{e} approximants may develop similar spurious singularities even at
high order and even in spatial regions where the approximants have started
to converge; the difficulty is not due to roundoff error, but rather to a
near-coincidence of zeros in the numerator and denominator of the rational
function which is the Pad\'{e} approximant \cite{B65,B75}. Similarly, the
three coefficients of the Hermite-Pad\' {e} quadratic and also the
discriminant, which is the argument of the square root in the
approximations, all have nearly coincident zeros as illustrated in Fig.~\ref
{FigTFMagHermitePadeKLM20_PQRDiscrim}. The branches also switch identities
at the singularities so that the physical solution is given by the ``plus''
branch on one side of the singularity and by the ``minus'' branch on the
other side.

This difficulty is not as serious as it seems. Both branches and their
average continue to give superb approximations to the eigenparameter $x_0 $.
The ordinary power series converges exponentially fast in the middle of the
spatial interval where the Hermite-Pad\'{e} fails.

The third surprise is that when the order of the approximant is thirty or
larger, both roots predict \emph{complex-valued} $x_{0}$. When $K=L=M=30$,
for example, the two branches of the quadratic give errors in $x_{0}$ of $%
0.199\times 10^{-23}+i0.166\times 10^{-15}$. The imaginary parts cancel when
the average is taken, leaving the extraordinarily small and real valued
average error of about $0.2\times 10^{-23}$! Thus, in spite of the three
surprises, which are also complications, and need for high precision
arithmetic at high order, the Hermit\'{e}-Pad\'e acceleration of the right
endpoint Taylor series is very successful.

\subsection{Power-series for a modified equation}

\begin{figure}[]
\centerline{\includegraphics[scale=1.0]{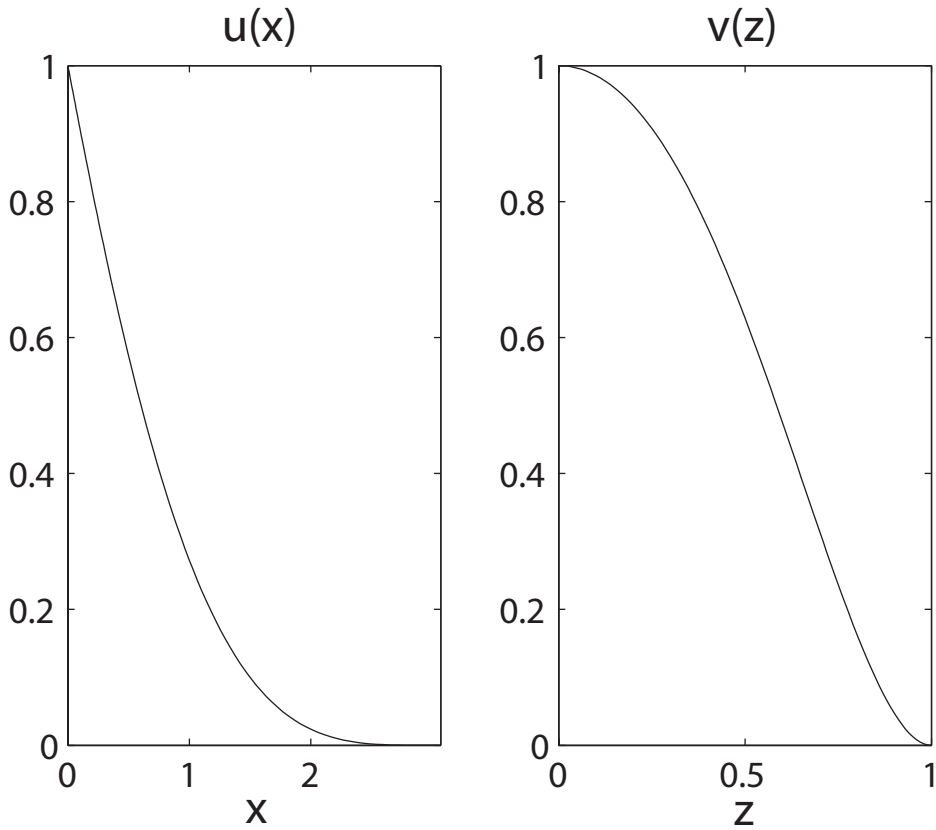}}
\caption{Left: the solution to the Thomas-Fermi problem or strong magnetic
field. Right: the transformed unknown $v (z) $ in the rescaled coordinate
space $z$.}
\label{FigTFMagnetic_u_and_v_plot}
\end{figure}

We can improve the results by removing the singularity through a convenient
transformation of the differential equation (\ref{eq:TFF_eq}). Instead of
the variables discussed above it is more convenient for our aims to choose $%
x=x_{0}z^{2}$ and $v(z)=\sqrt{u(x_{0}z^{2})}$ so that the differential
equation becomes
\begin{equation}
zv(z)v^{\prime \prime }(z)+zv^{\prime }(z)^{2}-v(z)v^{\prime
}(z)-2x_{0}^{5/2}z^{4}v(z)=0  \label{eq:TFF_eq_z}
\end{equation}
In this way the domain size $x_{0}$ appears as a sort of eigenparameter $%
x_{0}$. Fig~\ref{FigTFMagnetic_u_and_v_plot} compares the two functions $%
u(x) $ and $v(z)$.

We can thus obtain $x_{0}$ and $u_{0}^{\prime }$ from the partial sums
\begin{eqnarray}
v_{M}(z) &=&x_{0}^{5/2}t^{2}\sum_{j=0}^{M}\tilde{c}_{j}t^{j}  \nonumber \\
&=&x_{0}^{5/2}t^{2}\left( \frac{1}{3}+\frac{10}{21}t+\frac{149}{588}%
t^{2}+\ldots \right) ,\,t=z-1  \label{eq:v[M](z)}
\end{eqnarray}
and the equations $v_{M}(0)=1$ and $u_{0}^{\prime }=v_{M}^{\prime \prime
}(0)/x_{0}$. Results can be considerably improved by means of Pad\'{e}
approximants $[J,K](t)$. Here we choose diagonal $[M/2,M/2]$ and
near-diagonal $[(M-1)/2,(M+1)/2]$ ones because experience and theory suggest
that they are usually the most accurate\cite{B65,B75,BO78}. Tables \ref
{tab:p-s_sigma} and \ref{tab:p-s_xi} show the parameters $u_{0}^{\prime }$
and $x_{0}$ calculated by means of the partial sums (\ref{eq:v[M](z)}) and
their Pad\'{e} approximants. The rate of convergence is remarkable for a
power-series approach to a nonlinear problem.

\section{Chebyshev Pseudospectral Method}

\label{sec:Cheb_1}

We have also written a program that solves the Thomas-Fermi equation using
the Chebyshev pseudospectral method and Newton iteration. Because it is so
similar to an earlier for the Lane-Emden equation, we omit the details\cite
{B11}.

\subsection{Calculations of large order}

The solution is approximated in the form
\begin{equation}
v(z)=\sum_{n=0}^{\infty }a_{n}T_{n}(2z-1)
\end{equation}
The Chebyshev polynomials can be conveniently evaluated by $T_{n}(z)=\cos
(n\,\mbox{arccos}(z))$ or by the usual three term recurrence relation \cite
{B01}.

Newton's iteration, which was used to solve the system of
quadratic equations for the unknowns $(x_0 ,a_{0},a_{1},\ldots
a_{N})$, requires and initialization or ``first guess''. We found
that a first guess for $x_0 \in [2,6]$ combined with the lowest
term in the power series, $v\approx x_0 ^{5/2}(z-1)^{2}/3$,
suffice to give rapid convergence without underrelaxation.

\begin{figure}[]
\centerline{\includegraphics[scale=0.5]{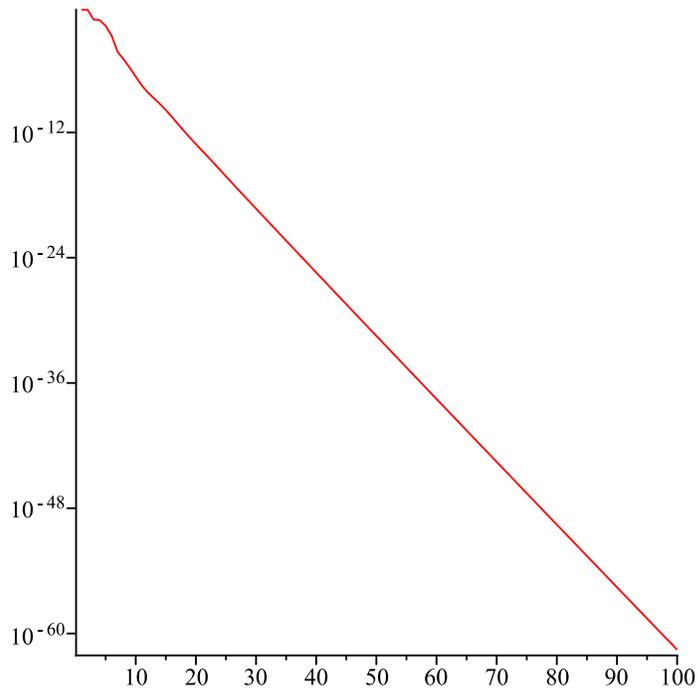}}
\caption{Chebyshev coefficients $a_{n}$ of $v(z)$.}
\label{FigTFMagChebcoef_N100}
\end{figure}

\begin{figure}[]
\centerline{\includegraphics[scale=0.5]{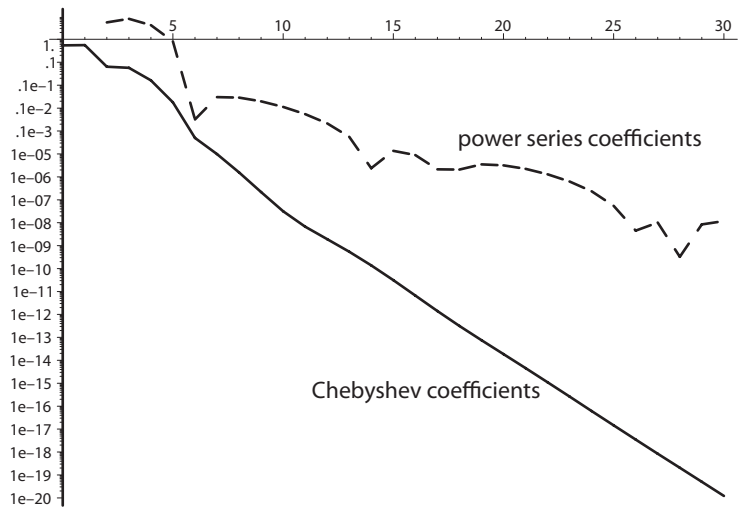}%
}
\caption{Chebyshev coefficients and power series coefficients of $v(z)$.}
\label{FigTFMagneticChebExpandMAPLEpowCheb}
\end{figure}

The Chebyshev coefficients converge geometrically as illustrated in Fig.~\ref
{FigTFMagChebcoef_N100} with $a_{n}\propto \exp (-n\mu )$ where $\mu $ is
between 1.4 and 1.42. All the Chebyshev coefficients larger than $10^{-15}$
listed in Table~\ref{TabCC}. The Chebyshev series converges much more
rapidly than the power series about the right endpoint as illustrated in
Fig.~\ref{FigTFMagneticChebExpandMAPLEpowCheb}.

Based on the trends in the Chebyshev calculations, we believe the
following results are accurate to all 50 decimal places shown.

\begin {equation}
x_0 = 3.06885718281479942624073100623167158584582595057745
\end{equation}

\begin {equation}
u^\prime_0 = -0.93896688764395889305505340187460180383289370739437
\end{equation}

\subsection{Semi-Analytical Solutions: Chebyshev Polynomial Methods for
Small $N$}

\label{sec:Cheb_2}

The most efficient spectral representation is one that incorporates all
three boundary conditions into the approximation in a form independent of
the spectral coefficients $d_{n}$:
\begin{equation}
v=(z-1)^{2}\,\left\{
1+d_{1}[T_{1}(2z-1)+T_{0}]+d_{2}[T_{2}(2z-1)+T_{1}(2z-1)+\ldots \right\}
\end{equation}
where we have used the identity $\mbox{sign}(T_{n}(-1))=-\mbox{sign}%
(T_{n+1}(-1))$ for all $n$. It is convenient, to eliminate fractional powers
in the unknowns, to define the new unknown
\begin{equation}
\lambda \equiv 2x_0 ^{5/2}
\end{equation}
The pseudospectral method with collocation at the interior points $%
x_{j}=(1+\cos (\pi j/(N+2))$, $j=1,\ldots (N+1)$ yields a system of coupled
polynomial equations, quadratic in the unknowns $(\lambda
,d_{1},d_{2},\ldots d_{N})$.

\begin{figure}[]
\centerline{\includegraphics[scale=1.0]{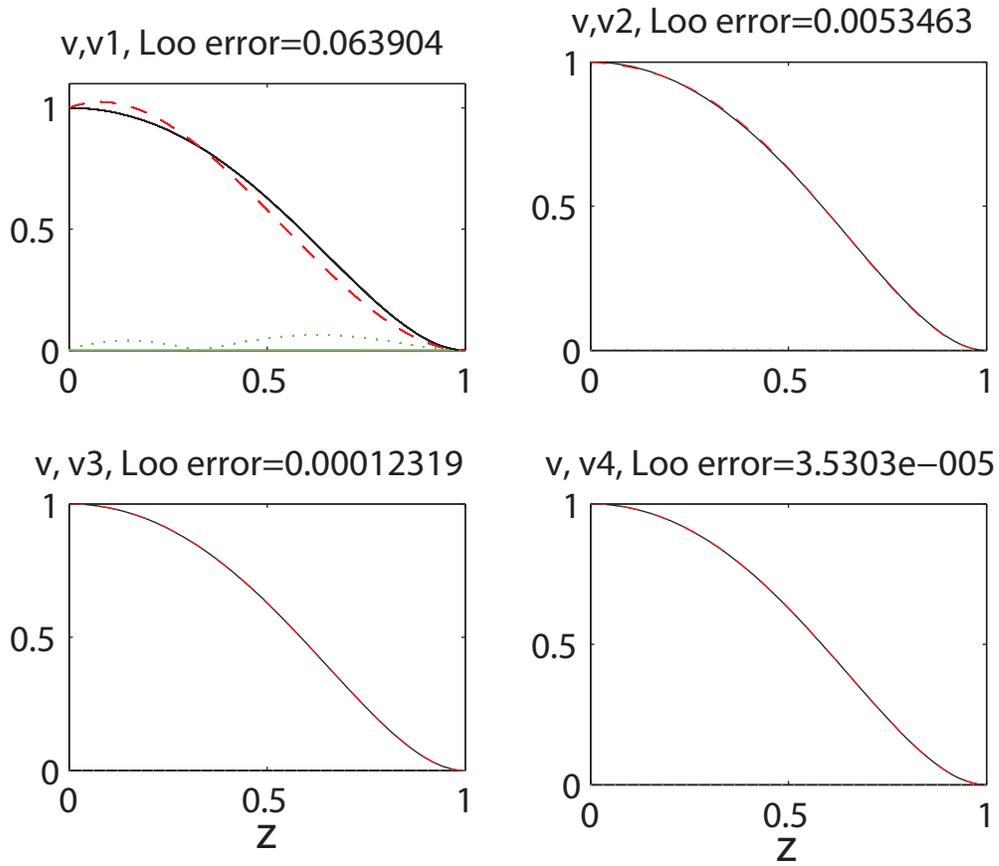}}
\caption{Comparisons of $v(z)$ [solid black] with $N=1$ to $N=4$ Chebsyhev
pseudospectral approximations [dashed red]. The green dotted lines, barely
visible at the bottom, are the errors.}
\label{FigTFMagnetic_smallNCheb_plot}
\end{figure}

For $N=1$, for example, substitution of $v=(z-1)^{2}\,(1+2d_{1}z)$ into the
differential equation yields the residual function
\begin{eqnarray}
resid(z;\lambda ,d_{1}) &=&z\,v\,v^{\prime \prime }+z(v^{\prime
})^{2}-\,v\,v^{\prime }-\lambda \,z^{4}\,v  \nonumber \\
&=&2-2\,\lambda \,z^{7}a_{{0}}+4\,\lambda
\,z^{6}a_{{0}}-2\,\lambda
\,z^{5}a_{{0}}-6\,z^{2}-64\,z^{3}a_{{0}}+36\,a_{{0}}z^{2} \nonumber \\
 &-&120\,z^{4}a_{{0}}^{2}
+96\,z^{3}a_{{0}}^{2}-24\,a_{{0}}^{2}z^{2}+30\,z^{4}a_{{0}}+48\,z^{5}a_{{0}%
}^{2}-\lambda \,z^{6}+2\,\lambda \,z^{5}\nonumber\\ &-&\lambda
\,z^{4}+4\,z^{3}-2\,a_{{0}}
\nonumber \\
\end{eqnarray}
The coupled system of two equations in two unknowns is the pair of equations
$resid(1/4;\lambda ,d_{1})=0$ and $resid(3/4;\lambda ,d_{1})=0$:
\begin{eqnarray}
{\frac{27}{16}}-{\frac{81}{128}}\,d_{{1}}-{\frac{27}{64}}\,{d_{{1}}}^{2}-{%
\frac{9}{4096}}\,\lambda -{\frac{9}{8192}}\,\lambda \,d_{{1}}=0 &&  \nonumber
\\
{\frac{5}{16}}+{\frac{95}{128}}\,d_{{1}}+{\frac{27}{64}}\,{d_{{1}}}^{2}-{%
\frac{81}{4096}}\,\lambda -{\frac{243}{8192}}\,\lambda \,d_{{1}} &=&0
\end{eqnarray}
The resultant of the two equations with elimination of $d_{1}$ is
\begin{equation}
{\frac{2705319}{4194304}}+{\frac{1643895}{268435456}}\,\lambda -{\frac{%
51182361}{68719476736}}\,{\lambda }^{2}+{\frac{1240029}{2199023255552}}\,{%
\lambda }^{3}=0
\end{equation}
This, with substitution of each $\lambda $ root in turn back into one of the
residual conditions to determine $d_{1}$, yields the three solutions:
\begin{eqnarray}
\{\lambda =34.33616,d_{1}=1.31544\};\{\lambda
=1311.8,d_{1}=-.66653\}\\ \nonumber \{\lambda
=-25.3931,d_{1}=-2.8724\}
\end{eqnarray}
The first solution is graphed in figure ~\ref{FigTFMagnetic_smallNCheb_plot}%
; the other two are spurious.

For small $N$, the Maple ``solve'' command will find all the finite
solutions to the system, here $2^{N+1}-1$ in number. In this work, the
physical solution was identified as that solution with the eigenparameter $%
\lambda $ closest to known values found by other means. When no such \emph{a
priori} information is available, a tedious but reliable procedure is to
substitute each solution into the differential equation to calculate the
residual, and accept the solution with the smallest residual norm. If
multiple solutions are suspected, a good strategy is to use each small-$N$%
solution as a first guess for the Newton pseudospectral code at
higher-resolution.

Maple's system solver is very slow and failed for $N=5$; in contrast, the
Newton/collocation program needed only 60 seconds to calculate for $N=100$
in 100 decimal place arithmetic. However, the Maple code that exploits the
''solve'' command is much shorter and is given in its entirety in Table~\ref
{TabMaplesmallC}.

The numerical results, including coefficients up to and including
$d_{N}$, are in Table~\ref{TabsmallN} and
Fig.~\ref{FigTFMagnetic_smallNCheb_plot}. Note that only the $N=1$
approximation, obtained by solving a pair of equations for
$(\lambda ,d_{1})$, is graphically distinguishable from the exact
$v(z)$. The success of such low truncations emphasizes the
smoothness of the Thomas-Fermi solution in the transformed
coordinate $z$ and unknown $u$. it also illustrates that spectral
methods can often be used in a semi-analytical mode as described
further in \cite{B93} and Chapter 20 of \cite{B01}.

\section{Conclusions}

\label{sec:conclusions}

We have shown that the PHM converges for two Thomas-Fermi equations that
exhibit quite different boundary conditions. In the case of neutral isolated
atoms we have $u(x\rightarrow \infty )=0$ and $u^{\prime }(x\rightarrow
\infty )=0$ while, on the other hand, $u(x_{0})=0$ and $u^{\prime }(x_{0})=0$
apply to the neutral atoms in very strong magnetic fields. If the success of
the PHM depends on the existence of a movable singularity that the approach
can push to infinity in the form of a zero of the denominator of the
Pad\'{e} approximant\cite{AB11}, then it seems that two different kinds of
singularities are involved in the problems just outlined. According to
Abbasbandy and Bervillier\cite{AB11}, in the former case such singular point
is precisely the pole $x_{s}$ (see Eq.~(\ref{eq:TF_pole})). However, in the
latter case the only candidate appears to be the zero/branch-point $x_{q}$
(see Eq.~(\ref{eq:TFF_zero})). Besides, in this case the rate of convergence
of the PHM is remarkably larger.

Present PHM calculations of the critical slope at the origin are quite
accurate but the generation of analytic Hankel determinants of dimension as
large as $D=60$ is time consuming. However, the rate of convergence is
commonly so great that one can obtain reasonably accurate results from
determinants of relatively small dimension. This feature of the PHM was
already exploited by Abbasbandy and Bervillier\cite{AB11} to estimate the
parameters of the conformal mapping used in the analytical continuation of
the power series.

We have also shown that nowadays available computer algebra
software like Mathematica enable us to obtain the solution to a
nonlinear differential equation quite efficiently and accurately
by means of suitable built in commands. Such results are shown in
tables \ref{tab:TFD}, \ref{tab:TFD} and \ref{tab:TFF}. However, we
found the PHM more convenient for the accurate calculation of the
slope at the origin as shown in tables \ref{tab:TF_crit} and
\ref{tab:TFF_crit}.

In the case of the Thomas-Fermi equation for a neutral atom in a very strong
magnetic field we have shown that the power-series expansion about the zero
of multiplicity two at $x_{0}$ is a suitable way of obtaining reasonably
accurate approximants to the solution over the entire physical interval $%
0<x<x_{0}$. The results can be slightly improved by means of Pad\'{e}
approximants and considerably improved by means of Hermite-Pad\'{e}
approximations, even when the latter exhibit a singular point or yield
complex results at large orders of approximation. If we remove the
singularity at origin by means of a suitable transformation both the power
series and its Pad\'{e} approximants lead to remarkably more accurate
results.

Finally, we have proved once more that the Chebyshev polynomials are by far
the most accurate and efficient way of solving this type of equations. In
this application we resorted to another computer-algebra software, Maple.

\ack

This work was supported by the National Science Foundation through
grant OCE 0451951

One of the authors (F.M.F) wants to thank Professor C. Bervillier
for useful comments and suggestions

\begin{table}[]
\caption{Critical slope at origin for the Thomas-Fermi equation for neutral
atoms}
\label{tab:TF_crit}
\begin{center}
\begin{tabular}{|l|l|}
\hline Source & \multicolumn{1}{c|}{$u^\prime_0$} \\ \hline
Ref.\cite{KMNU55} & -1.588071 \\
Ref.\cite{T72} & -1.5880710 \\
Ref.\cite{F11} & -1.588071022611375313 \\
Ref.\cite{AB11} & -1.5880710226113753127189$\pm 7\times 10^{-22}$ \\
Ref.\cite{B12} & -1.5880710226113753127186845 \\
Present (integration) & -1.588071022611375312718684 \\
Present (PHM) & -1.588071022611375312718684508 \\ \hline
\end{tabular}
\end{center}
\end{table}

\begin{table}[]
\caption{Critical slope at origin for the Thomas-Fermi equation for neutral
atoms in strong magnetic fields}
\label{tab:TFF_crit}
\begin{center}
\begin{tabular}{|l|l|}
\hline Source & \multicolumn{1}{c|}{$u^\prime_0$} \\ \hline
Ref.\cite{BCR74} & -0.93896594 \\
Ref.\cite{HGM83} & -0.938966887644 \\
Ref.\cite{F11} & -0.93896688764395889306 \\
Present (PHM) & -0.9389668876439588930550534018746018038328937073944 \\
\hline
\end{tabular}
\end{center}
\end{table}

\begin{table}[]
\caption{Solution to the Thomas-Fermi equation for neutral atoms}
\label{tab:TF}
\begin{center}
{\tiny
\begin{tabular}{|r|l|}
\hline \multicolumn{1}{|c|}{$x$} & \multicolumn{1}{c|}{$u(x)$} \\
\hline
0 &   1.\\
10 &    0.024314292988681\\
20 &    0.0057849411915669\\
30 &    0.0022558366162029\\
40 &    0.0011136356388334\\
50 &    0.0006322547829849\\
60 &    0.00039391136668542\\
70 &    0.00026226529981201\\
80 &    0.00018354575974071\\
90 &    0.00013354582895373\\
100 &   0.00010024256813941\\
110 &   0.000077192183914122\\
120 &   0.000060724454048042\\
130 &   0.000048642170611589\\
140 &   0.000039574139448193\\
150 &   0.000032633964446257\\
160 &   0.000027231036946158\\
170 &   0.000022961351007399\\
180 &   0.000019542101672266\\
190 &   0.000016771248041131\\
200 &   0.000014501803496946\\
210 &   0.00001262507868731\\
220 &   0.000011059514500269\\
230 &   9.7430900509356e-6\\
240 &   8.6280669794388e-6\\
250 &   7.6772907646176e-6\\
260 &   6.8615483807441e-6\\
270 &   6.1576544141983e-6\\
280 &   5.5470471166532e-6\\
290 &   5.0147463894448e-6\\
300 &   4.548571953617e-6\\
310 &   4.1385507917383e-6\\
320 &   3.7764638007387e-6\\
330 &   3.4554958931053e-6\\
340 &   3.1699637128939e-6\\
350 &   2.9151021105849e-6\\
360 &   2.6868954790918e-6\\
370 &   2.4819436135582e-6\\
380 &   2.2973543393218e-6\\
390 &   2.1306570419327e-6\\
400 &   1.9797326281136e-6\\
410 &   1.842756484985e-6\\
420 &   1.7181517839461e-6\\
430 &   1.6045510644226e-6\\
440 &   1.5007644808624e-6\\
450 &   1.4057534397658e-6\\
460 &   1.3186086183364e-6\\
470 &   1.2385315617813e-6\\
480 &   1.1648192165876e-6\\
490 &   1.0968508828835e-6\\
500 &   1.034077168203e-6\\
510 &   9.7601060362214e-7\\
520 &   9.2221764589762e-7\\
530 &   8.7231183937678e-7\\
540 &   8.2594795176694e-7\\
550 &   7.828169303947e-7\\
560 &   7.4264155197267e-7\\
570 &   7.0517266036529e-7\\
580 &   6.7018590439059e-7\\
590 &   6.3747890208209e-7\\
600 &   6.0686876967458e-7\\
610 &   5.7818996335412e-7\\
620 &   5.5129238991189e-7\\
630 &   5.2603974917333e-7\\
640 &   5.0230807668597e-7\\
650 &   4.7998445984162e-7\\
660 &   4.5896590454401e-7\\
670 &   4.3915833284169e-7\\
680 &   4.2047569473637e-7\\
690 &   4.0283917973564e-7\\
700 &   3.861765157183e-7\\
710 &   3.7042134437901e-7\\
720 &   3.5551266396608e-7\\
730 &   3.4139433126073e-7\\
740 &   3.2801461580316e-7\\
750 &   3.1532580027683e-7\\
760 &   3.032838217405e-7\\
770 &   2.918479490687e-7\\
780 &   2.8098049253898e-7\\
790 &   2.7064654200495e-7\\
800 &   2.6081373052692e-7\\
810 &   2.5145202070801e-7\\
820 &   2.4253351131026e-7\\
830 &   2.3403226200994e-7\\
840 &   2.2592413439939e-7\\
850 &   2.1818664755992e-7\\
860 &   2.1079884671993e-7\\
870 &   2.0374118367901e-7\\
880 &   1.9699540782507e-7\\
890 &   1.9054446669994e-7\\
900 &   1.8437241518212e-7\\
910 &   1.7846433245545e-7\\
920 &   1.7280624602028e-7\\
930 &   1.6738506208215e-7\\
940 &   1.6218850172182e-7\\
950 &   1.5720504231184e-7\\
960 &   1.5242386369939e-7\\
970 &   1.4783479872324e-7\\
980 &   1.4342828767613e-7\\
990 &   1.3919533636191e-7\\
1000 &   1.3512747743129e-7\\

\hline
\end{tabular}
}
\end{center}
\end{table}

\begin{table}[]
\caption{First derivative of the Solution to the Thomas-Fermi equation for
neutral atoms}
\label{tab:TFD}
\begin{center}
{\tiny
\begin{tabular}{|r|l|}
\hline \multicolumn{1}{|c|}{$x$} &
\multicolumn{1}{c|}{$u^{\prime}(x)$} \\
\hline 0 &   -1.5880710226114\\
10 &   -0.0046028818712693\\
20 &   -0.00064725433277769\\
30 &   -0.00018067000647699\\
40 &   -0.000069668028540326\\
50 &   -0.000032498902048259\\
60 &   -0.0000171977000831\\
70 &   -9.9565333930524e-6\\
80 &   -6.1661955287641e-6\\
90 &   -4.0244737037667e-6\\
100 &   -2.7393510686783e-6\\
110 &   -1.9299022690995e-6\\
120 &   -1.3992583170223e-6\\
130 &   -1.0395157118193e-6\\
140 &   -7.8856447021428e-7\\
150 &   -6.0913994786089e-7\\
160 &   -4.7807415548727e-7\\
170 &   -3.8051134288369e-7\\
180 &   -3.0666432312939e-7\\
190 &   -2.4992901331678e-7\\
200 &   -2.0575323164753e-7\\
210 &   -1.7093867858999e-7\\
220 &   -1.4319925950792e-7\\
230 &   -1.2087522043829e-7\\
240 &   -1.0274430764511e-7\\
250 &   -8.7894679831444e-8\\
260 &   -7.5637914908074e-8\\
270 &   -6.544852781645e-8\\
280 &   -5.6921313455898e-8\\
290 &   -4.9740860705699e-8\\
300 &   -4.3659496185299e-8\\
310 &   -3.8481144114006e-8\\
320 &   -3.4049389482697e-8\\
330 &   -3.0238562022714e-8\\
340 &   -2.6947014494947e-8\\
350 &   -2.4092011004349e-8\\
360 &   -2.1605807799308e-8\\
370 &   -1.9432625152504e-8\\
380 &   -1.7526290677749e-8\\
390 &   -1.5848392577797e-8\\
400 &   -1.4366823059949e-8\\
410 &   -1.3054622396184e-8\\
420 &   -1.1889056199565e-8\\
430 &   -1.0850874763838e-8\\
440 &   -9.9237153936954e-9\\
450 &   -9.0936176860656e-9\\
460 &   -8.3486285238475e-9\\
470 &   -7.6784786982998e-9\\
480 &   -7.0743170080134e-9\\
490 &   -6.5284906994362e-9\\
500 &   -6.0343634424472e-9\\
510 &   -5.5861638415855e-9\\
520 &   -5.1788588934186e-9\\
530 &   -4.8080479060999e-9\\
540 &   -4.4698732683358e-9\\
550 &   -4.160945144665e-9\\
560 &   -3.878277722421e-9\\
570 &   -3.6192350738087e-9\\
580 &   -3.3814850478544e-9\\
590 &   -3.1629598899054e-9\\
600 &   -2.9618225150474e-9\\
610 &   -2.7764375473743e-9\\
620 &   -2.6053463881518e-9\\
630 &   -2.4472456993964e-9\\
640 &   -2.3009687906329e-9\\
650 &   -2.1654694798767e-9\\
660 &   -2.0398080686052e-9\\
670 &   -1.9231391273669e-9\\
680 &   -1.8147008358921e-9\\
690 &   -1.7138056608835e-9\\
700 &   -1.6198321874803e-9\\
710 &   -1.5322179478635e-9\\
720 &   -1.4504531135232e-9\\
730 &   -1.3740749371114e-9\\
740 &   -1.3026628461653e-9\\
750 &   -1.2358341048206e-9\\
760 &   -1.1732399713605e-9\\
770 &   -1.1145622894042e-9\\
780 &   -1.0595104590168e-9\\
790 &   -1.0078187412534e-9\\
800 &   -9.59243855836e-10\\
810 &   -9.135628369537e-10\\
820 &   -8.7057111672569e-10\\
830 &   -8.3008080977438e-10\\
840 &   -7.9191917572404e-10\\
850 &   -7.5592723934792e-10\\
860 &   -7.2195855060047e-10\\
870 &   -6.8987806894922e-10\\
880 &   -6.5956115831007e-10\\
890 &   -6.3089268053242e-10\\
900 &   -6.0376617681006e-10\\
910 &   -5.7808312764064e-10\\
920 &   -5.5375228304499e-10\\
930 &   -5.3068905571025e-10\\
940 &   -5.0881497055441e-10\\
950 &   -4.8805716494193e-10\\
960 &   -4.6834793442266e-10\\
970 &   -4.4962431943178e-10\\
980 &   -4.3182772888659e-10\\
990 &   -4.1490359705517e-10\\
1000 &   -3.9880107046013e-10\\
\hline
\end{tabular}
}
\end{center}
\end{table}

\begin{table}[]
\caption{Solution to the Thomas-Fermi equation for neutral atoms in strong
magnetic fields}
\label{tab:TFF}
\begin{center}
{\tiny
\begin{tabular}{|r|l|}
\hline \multicolumn{1}{|c|}{$x$} & \multicolumn{1}{c|}{$u(x)$} \\
\hline
0. &   1.\\
0.1 &   0.7170931035374508\\
0.2 &   0.6124516884444692\\
0.3 &   0.5381961080581211\\
0.4 &   0.479844885457407\\
0.5 &   0.4317097837447453\\
0.6 &   0.3908409649565394\\
0.7 &   0.3554696879373627\\
0.8 &   0.3244342852150042\\
0.9 &   0.2969222776883283\\
1. &   0.2723385646065789\\
1.1 &   0.2502315742397729\\
1.2 &   0.2302489290264509\\
1.3 &   0.2121093520054812\\
1.4 &   0.1955840735228681\\
1.5 &   0.1804840769457186\\
1.6 &   0.1666510831170969\\
1.7 &   0.1539510127378397\\
1.8 &   0.1422691401494945\\
1.9 &   0.1315064314341758\\
2. &   0.1215767304319089\\
2.1 &   0.1124045638520199\\
2.2 &   0.1039234063491158\\
2.3 &   0.09607429270074567\\
2.4 &   0.08880469561358382\\
2.5 &   0.0820676094016386\\
2.6 &   0.07582079507183087\\
2.7 &   0.07002615329387998\\
2.8 &   0.06464919967550395\\
2.9 &   0.05965862260930963\\
3. &   0.05502590831205111\\
3.1 &   0.05072502095749841\\
3.2 &   0.04673212830182475\\
3.3 &   0.04302536512062993\\
3.4 &   0.0395846282664133\\
3.5 &   0.03639139832082583\\
3.6 &   0.03342858373515079\\
3.7 &   0.0306803840826662\\
3.8 &   0.02813216963071108\\
3.9 &   0.02577037491068775\\
4. &   0.02358240434538228\\
4.1 &   0.02155654830361774\\
4.2 &   0.01968190820681658\\
4.3 &   0.01794832952175207\\
4.4 &   0.0163463416473783\\
4.5 &   0.01486710384803269\\
4.6 &   0.01350235650595824\\
4.7 &   0.0122443770673297\\
4.8 &   0.01108594014126138\\
4.9 &   0.01002028128341276\\
5. &   0.00904106405704823\\
5.1 &   0.008142350016579636\\
5.2 &   0.007318571303218951\\
5.3 &   0.006564505580616747\\
5.4 &   0.005875253071267334\\
5.5 &   0.005246215482854354\\
5.6 &   0.004673076638280921\\
5.7 &   0.004151784644449602\\
5.8 &   0.003678535453408486\\
5.9 &   0.003249757685661876\\
6. &   0.002862098599594876\\
\hline
\end{tabular}
}
\end{center}
\end{table}

\begin{table}[tbp]
\caption{Errors in Eigenlength $x_0$ for the magnetic case, obtained from
right endpoint power series and Hermite-Pad\'{e} approximation for $u(x)$}
\label{Tab1eHermPad}
\begin{tabular}{|r|r|l|l|l|l|}
\hline
$K$ ($=L=M$) & $N$ & Taylor series & first root & second root & average of
roots \\ \hline
2 & 7 & 0.0039 & 0.0023 & 0.044 & 0.023 \\
5 & 16 & 0.00076 & 0.00093 & 0.000052 & 0.00043 \\
10 & 31 & $0.174\times 10^{-3}$ & $0.265\times 10^{-6}$ & $0.32\times
10^{-6} $ & $0.31\times 10^{-7}$ \\
15 & 46 & $0.69\times 10^{-4}$ & $0.11\times 10^{-8}$ & $0.11\times 10^{-8}$
& $0.32\times 10^{-11}$ \\
20 & 61 & $0.35\times 10^{-4}$ & $0.15\times 10^{-11}$ & $0.15\times
10^{-11} $ & $0.45\times 10^{-15}$ \\
25 & 76 & $0.21\times 10^{-4}$ & $0.11\times 10^{-12}$ & $0.11\times
10^{-12} $ & $0.39\times 10^{-18}$ \\ \hline
\end{tabular}
\end{table}

\begin{table}[]
\caption{$v$-power series approximation of initial slope
$u^\prime_0$; correct digits in boldface} \label{tab:p-s_sigma}
\begin{tabular}{|c|c|l|l|}
\hline $M$ & Method & $u^\prime_0$ & Number of correct digits \\
\hline
10 & power series & \textbf{-0.93}64 & 2 \\
& Pad\'e & \textbf{-0.93}27 & 2 \\
20 & power series & \textbf{-0.938}79 & 3 \\
& Pad\'e & \textbf{-0.93896}45 & 5 \\
40 & power series & \textbf{-0.9389668}34 & 7 \\
& Pad\'e & \textbf{-0.938966887}60 & 9 \\
60 & power series & \textbf{-0.9389668876}35 & 10 \\
& Pad\'e & \textbf{-0.93896688764395}54 & 14 \\
80 & power series & \textbf{-0.93896688764395}74 & 14 \\
& Pad\'e & \textbf{-0.93896688764395889}27 & 17 \\
--- & PHM\cite{F11} & \textbf{-0.93896688764395889306} & 20 \\ \hline
\end{tabular}
\end{table}

\begin{table}[]
\caption{Eigenparameter $x_0$ approximations from the $v$-power series}
\label{tab:p-s_xi}
\begin{tabular}{|c|c|l|}
\hline
$M$ & Method & $x_0$ \\ \hline
10 & power series & \textbf{3.0688}06 \\
& Pad\'e & \textbf{3.068}77 \\
20 & power series & \textbf{3.06885}60 \\
& Pad\'e & \textbf{3.0688571}76 \\
40 & power series & \textbf{3.068857182}71 \\
& Pad\'e & \textbf{3.0688571828147994}51 \\
60 & power series & \textbf{3.06885718281479}17 \\
& Pad\'e & \textbf{3.068857182814799426240731}39 \\
80 & power series & \textbf{3.06885718281479942}55 \\
& Pad\'e & \textbf{3.068857182814799426240731006231672626130} \\ \hline
\end{tabular}
\end{table}

\begin{table}[]
\caption{Chebyshev coefficients of the solution to the magnetic case of the
Thomas-Fermi equation}
\label{TabCC}
\begin{tabular}{|c|c|}
\hline
degree $n$ & $a_{n}$ \\ \hline
0 & $5.48135788388634344\times 10{-01}$ \\
1 & $-5.58881613523477153\times 10{-01}$ \\
2 & $-6.42705283349619230\times 10{-02}$ \\
3 & $5.70880541499120995\times 10{-02}$ \\
4 & $1.61861107728602653\times 10{-02}$ \\
5 & $1.78344934649533460\times 10{-03}$ \\
6 & $-4.97638953814069498\times 10{-05}$ \\
7 & $9.88568187276911699\times 10{-06}$ \\
8 & $-1.57314041941070364\times 10{-06}$ \\
9 & $2.17031601079801803\times 10{-07}$ \\
10 & $-3.17277060187282838\times 10{-08}$ \\
11 & $6.73733273124207577\times 10{-09}$ \\
12 & $-1.91849909540107163\times 10{-09}$ \\
13 & $5.43530178894964329\times 10{-10}$ \\
14 & $-1.37539658139417754\times 10{-10}$ \\
15 & $3.12411846416475022\times 10{-11}$ \\
16 & $-6.65185565734159168\times 10{-12}$ \\
17 & $1.41167490452475800\times 10{-12}$ \\
18 & $-3.15485896233555226\times 10{-13}$ \\
19 & $7.52433060541288362\times 10{-14}$ \\
20 & $-1.85848288796828985\times 10{-14}$ \\
21 & $4.57976772880843444\times 10{-15}$ \\
22 & $-1.10484891112084068\times 10{-15}$ \\ \hline
\end{tabular}
\end{table}

\begin{table}[h]
\caption{Maple Code to Compute a Chebyshev Pseudospectral Approximation for
Small $N$}
\label{TabMaplesmallC}
\begin{tabular}{|l|}
\hline
restart; with(orthopoly); N:=4; \# Define approximation in next 2 lines; \\
vv:=1; for n from 1 to N do vv:=vv + d[n]*(T(n,2*x-1)+T(n-1,2*x-1)); od: \\
v:= (x-1)*(x-1)*vv; \# derivatives of v; vx:=diff(v,x); vxx:=diff(v,x,x); \\
resid:= x*v*vxx + x*vx*vx -v*vx - lambda* x**4 *v; \# ODE residual; \\
ta[0]:=evalf( Pi*(0+1)/(N+2) ); xa[0]:=evalf((cos(ta[0])+1)/2); \# Chebyshev
grid points "xa"; \\
resida[0]:= evalf(subs(x=xa[0],resid)); \\
varset:= {lambda}; eqset:= {resida[0]}; \# initialize set of unknowns \& the
set of pointwise residuals; \\
for j from 1 to N do ta[j]:= evalf( Pi*(j+1)/(N+2) ); \\
xa[j]:= evalf( (cos(ta[j] ) + 1)/2 ); resida[j]:=
evalf(subs(x=xa[j],resid)); \# array of ODE residuals ; \\
varset:= varset union {d[j]}; eqset:= eqset union {resida[j]}; \# update
variable \& equation sets; od: \\
solutionvector:=solve(eqset,varset); assign(solutionvector[1]); \# solve the
polynomial system; \\
\# \{resida[j]=0, j=0, ..., N\} in unknowns \{$\lambda, d_{1}, \ldots d_{N} $%
\} ; \\
xi:= evalf( (lambda/2)**(2/5) ); sigma:= evalf( subs(x=0,diff(v,x,x)/xi) );
\\ \hline
\end{tabular}
\end{table}

\begin{table}[h]
\caption{Results and Errors from Low Order Collocation}
\label{TabsmallN}
\begin{tabular}{|c|c|c|c|c|c|}
\hline
Name & $N=1$ & $N=2$ & $N=3$ & $N=4$ & Exact \\ \hline
$x_0$ & 3.11809 & 3.07417 & 3.0687055 & 3.068876456 & 3.0688571 \\
$x_0$ error & 0.049 & 0.0053 & -0.00015 & $0.19274\times 10{-4}$ & --- \\
$u^\prime_0$ & -2.73 & -.281 & -.89047 & -.9237607 & -0.9389669 \\
$u^\prime_0$ error & -1.79 & 0.66 & 0.048 & 0.015 & --- \\
$d_{1}$ & 1.315 & 1.8211 & 1.8837159 & 1.882043824 & 1.8825071389 \\
$d_{2}$ &  & .29256 & .34172159 & .3402298497 & .34061038 \\
$d_{3}$ &  &  & 0.027764 & 0.0266875 & .0269894965 \\
$d_{4}$ &  &  &  & -0.000662 & -.0004613479537 \\ \hline
\end{tabular}
\end{table}

\clearpage

\clearpage

\end{document}